\documentclass[final]{svjour3}

\usepackage[margin=1.5in]{geometry}

\usepackage{graphicx}
\usepackage{rotating}
\usepackage{amsmath}
\usepackage{amssymb}
\usepackage{mathptmx}
\usepackage[numbers]{natbib}
\usepackage{color}
\usepackage{hyperref}  
\usepackage{comment}
\makeatletter
\journalname{Journal of Low Temperature Physics}

\begin{document}
\newcommand{\hdblarrow}{H\makebox[0.9ex][l]{$\downdownarrows$}-}
\title{Simulation and Measurement of Out-of-Band Resonances for the FDM Readout of a TES Bolometer}
\author{
A. Aminaei$^{1}$  H. Akamatsu$^{2}$  A.C.T. Nieuwenhuizen$^{2}$  D. Vaccaro$^{2}$  Q. Wang$^{2}$ M.D. Audley$^{2}$ P. Khosropanah$^{2}$ 
A. McCalden$^{2}$ D. Boersma$^{2}$ M. Ridder$^{2}$ S. Ilyas$^{3}$ J. van der Kuur$^{2}$ G. de Lange$^{2}$ }
\institute{1: University of California, Davis, USA.\\ 2: SRON Netherlands Institute for Space Research, Leiden/Groningen, the Netherlands.\\
3: ASML, Veldhoven, The Netherlands.\\ \email{aaminaei@ucdavis.edu}}
 \titlerunning{Simulation and Measurement of Out-of-Band Resonances for the FDM Readout of a TES Bolometer}
\authorrunning{Aminaei et al. }
\maketitle
\begin{abstract}
 \enspace With applications in cosmology, infrared astronomy and CMB survey, frequency-division multiplexing (FDM) proved to be a viable readout for transition-edge sensors (TESs). We investigate the occurrence of out-of-band resonances (OBR) which could constrain the bandwidth of the FDM readout of TES bolometers. The study includes SPICE modeling of the entire setup including the cryogenic harness, LC filters, Superconducting Quantum Interference Device (SQUID) and room-temperature amplifier. Simulation results show that the long harness (for flight model) could cause multiple reflections that generate repetitive spikes in the spectrum. Peaks of the OBR are mainly due to the parasitic capacitances at the input  of SQUID. Implementing a low-pass RC circuit (snubber) at the input of the SQUID dampened the OBR. As a result, the first peak only appears around 20 MHz which is a safe margin for the 1 MHz-3.8 MHz FDM in use in the prototype readout. Using a spectrum analyzer and broadband LNAs, we also measured the OBR for the prototype FDM readout in the lab up to 500 MHz.  The measurement was conducted at temperatures of 50 mK and 4 K and for various biasing of the DC SQUID. It turns out that OBR are more intense at 50 mK and are caused by the harness impedance mismatch rather than the SQUID. Simulation codes and supporting materials are available at \footnote[4]{https://github.com/githubamin/LT-Spice-Simulation-of-FDM-readout}.


\end{abstract}
\section{Introduction}
\begin{figure}[h]
\includegraphics[width=13.5cm]{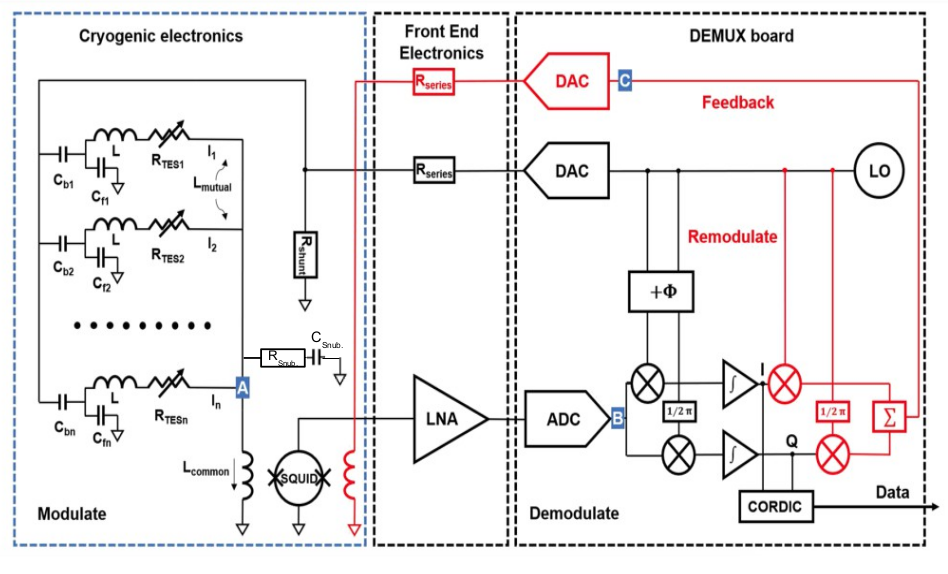}
\caption{A schematic of the BBFB FDM readout developed at SRON. Position of RC circuit (snubber) is identified at the summing point A. Two-stage SQUID has been used for both measurement and simulation. Reproduced after \cite{ref10}.  }
\label{fig01}
\end{figure}

FDM is a readout technique for the TES-based bolometer and (micro-)calorimeter arrays  
which are used in infrared, Cosmic Microwave Background, and X-ray astronomy \cite{infra1,infra2} \cite{cmb1, cmb2,cmb3} \cite{e-squid, ref01, ref02}.
 In the FDM readout, TES pixels are connected in series to high-Q LC resonators with the resonance frequencies within few MHz range and Q factor in the order of $10^4$ and higher. Signals from the individual TESs are summed up and read out by a DC SQUID amplifier in a sub-Kelvin cryogenic environment. At next stage, signals are amplified by a room temperature RF Low Noise Amplifier (LNA), then digitized and demodulated. 
 FDM readout is particularly considered for space applications due to its low power dissipation at cryogenic temperatures and light-weight shielding requirement. In addition, 
 TES operation points can be individually adjusted which identify the requirements of the SQUID specs \cite{ref01}. At  SRON, FDM readouts are developed with a baseband feedback (BBFB) module to increase the linear dynamic range of the system \cite{ref02}.  A schematic of the BBFB FDM readout developed at SRON is illustrated in Fig.~\ref{fig01}.  Cryogenic electronics include TES sensors, LC resonators and SQUID amplifiers. The SQUID in use for this study is a two-stage SQUID which has been also simulated accordingly. The FM modulated signal is summed up at point A and is amplified by the SQUID and then by LNA at the front-end electronics. 
 The amplified signal is digitized by the ADC (Analog to Digital Converter, point B) and is demodulated by the DEMUX board.  Using a DAC (Digital to Analog Converter, point C), an analog feedback is sent back to the feedback coil to enhance the linear performance of the SQUID. 

In Fig.\ref{fig01}, the $R_{shunt}$ is in parallel to the LC resonators and provides the AC bias using the AC bias circuit(not shown). Details of the BBFB FDM readout can be found in \cite{ref02}, \cite{ref10}. As we discuss in the following sections, the RC circuit (snubber) placed at the summing point (Point A) dampens the out of band resonances(OBR) which could cause a negative impact on the FDM readout. The use of snubber circuit improves the SQUID performance with smoother $V-\rm\Phi$ curves and lower noise \cite{ref31}, \cite{ref32}. In another FDM readout, it has been also reported that shunting only a 5 $\rm\Omega$ resistor with the input coil of the SQUID removes the SQUID irregularities. This is likely due to the fact that the shunt resistor and the input coil of the SQUID make an RL low-pass filter\cite{ref33}.   
\section{Simulation}
\begin{figure}[h]
\includegraphics[width=13.5cm]{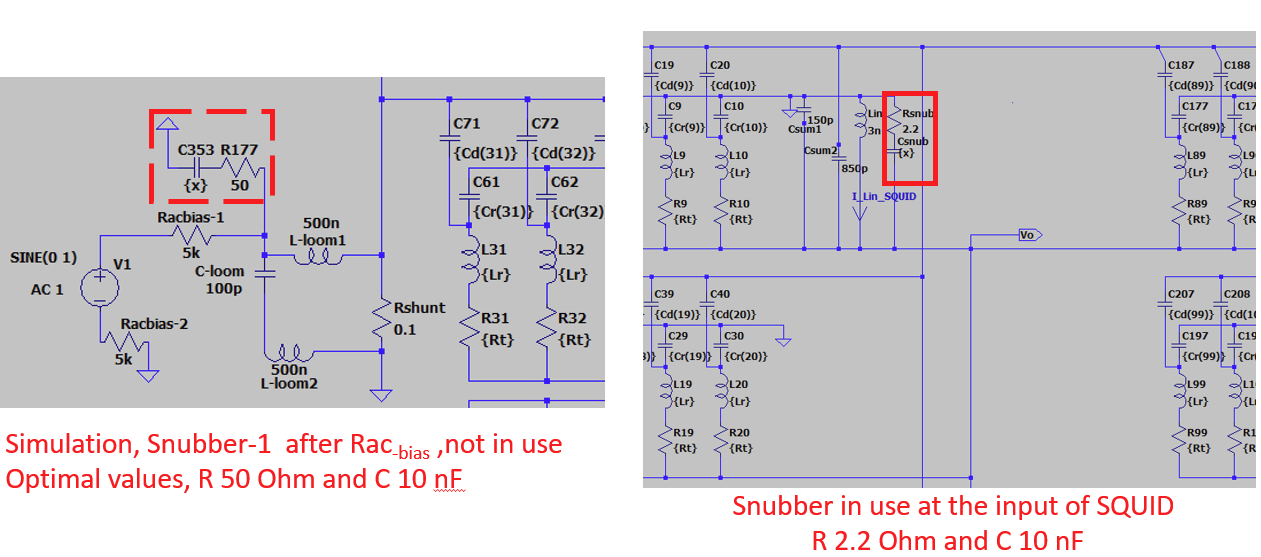}
\caption{An illustration of the impact of the cryogenic harness and input coil of the SQUID at Summing Point for the OBR of the FDM readout. Position of RC circuit (snubber) in use in the FDM prototype is shown in the right red solid box. The 2nd snubber (in left in dotted red box) is only used in simulation.}
\label{fig1}
\end{figure}

\begin{figure}[h]
\includegraphics[width=13.5cm]{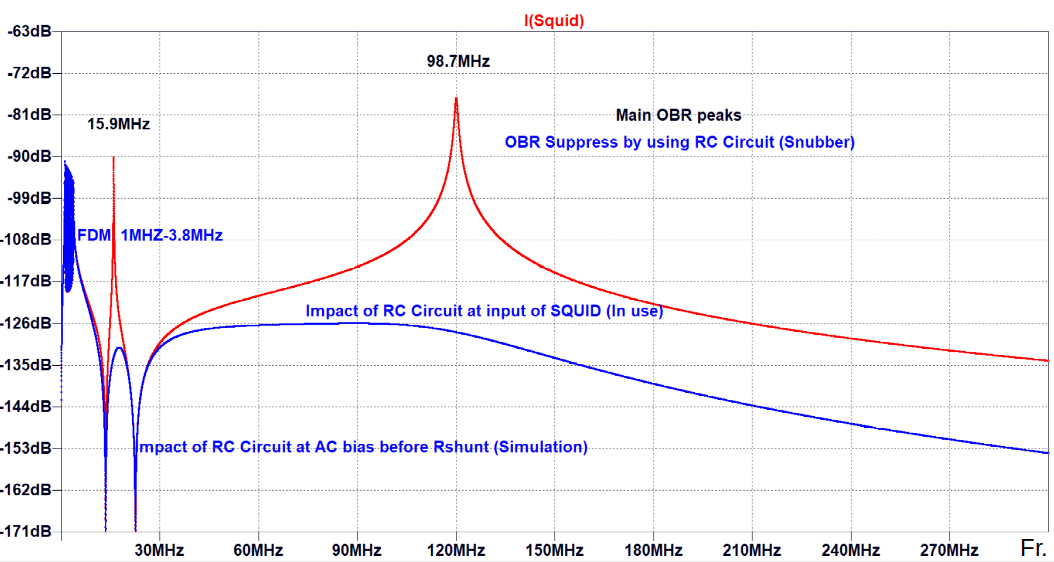}
\caption{\label{fig2}
Simulation results of Fig.~\ref{fig1}: variation of the intensity of input current of SQUID vs. frequency - with (Blue) and without (Red) snubbers(RC filter circuits). The peak at 15.9 MHz is due to the harness coupling and the peak at 98.7 MHz is due to the parasitic capacitances at the input of SQUID.
}
\end{figure}

\begin{figure}[h]
\includegraphics[width=13.5cm]{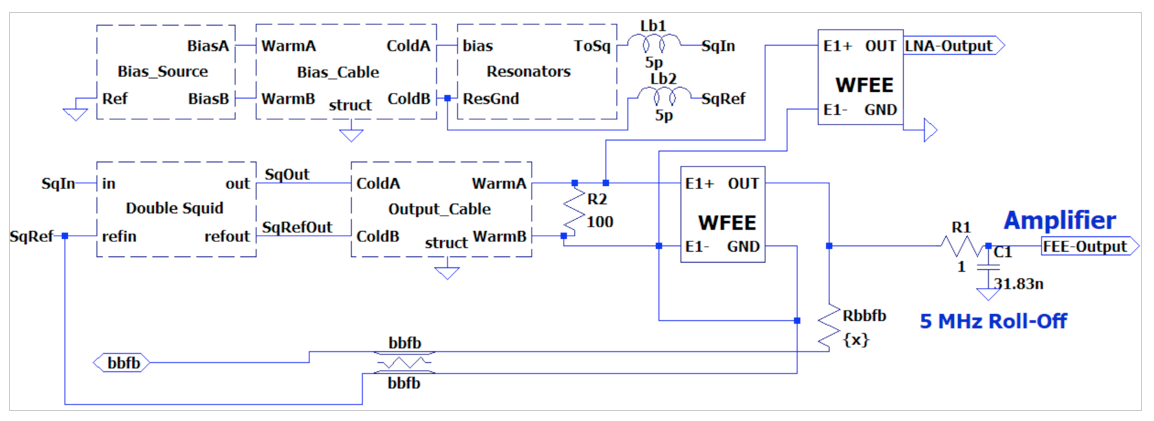}
\caption{\label{fig3}
Block diagram of the FDM readout of TES sensors. Blocks of AC bias voltage, input harness, LC resonators, double-stage SQUID and output harness are simulated. WFEE stands for Warm Front End Electronics. Baseband feedback is modeled with a variable resistor and transmission line. 
}
\end{figure}

\begin{figure}[h]
\includegraphics[width=13.5cm]{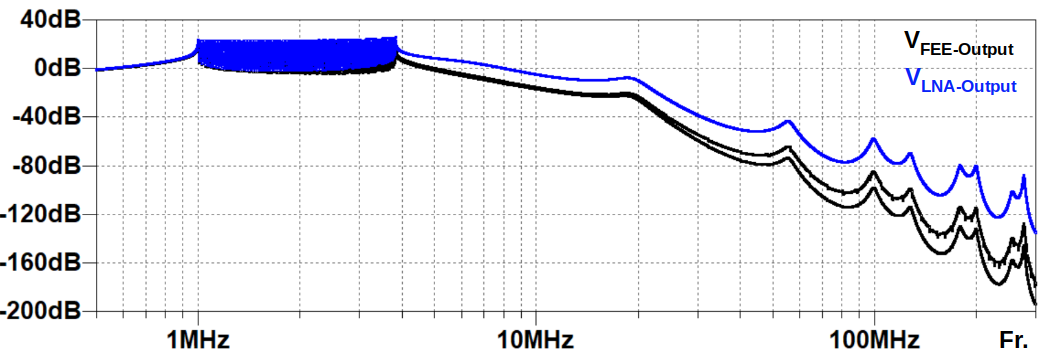}
\caption{\label{fig4}
Simulation results of the BBFB FDM circuit with a snubber(RC filter) at the input of SQUID (see Fig.~\ref{fig3} and Fig.~\ref{fig02}). Variation of intensity of output voltages vs. frequency: using the external LNA (blue) and FEE (black) with  $R_{\rm bbfb}$ 1 K$\rm\Omega$(top) and 10 K$\rm\Omega$.
}
\end{figure}

 In the FDM readout, TES information is embedded only in the FDM band so it is desirable to suppress the OBR as much as possible. The fundamentals of the baseband feedback(BBFB) FDM readout of TES detectors and an overview of the SRON FDM readout have been described in \cite{infra2}, \cite{ref2}. As can be seen in Fig. \ref{fig1}, the AC bias line, cryogenic harness, 176 LC resonators and input coil of the first SQUID (3 nH) are simulated using LT-SPICE software \cite{refSpice}. An AC bias reference of 1 V is provided via the AC bias line including 5 k$\rm\Omega$ resistors and cryogenic harness which is a lossless transmission line with characteristic impedance of 70.7 $\rm\Omega$(L=500 nH and C=100 pF). TES are assumed to be in the transition region \cite{ref5} with a resistance of 40 m$\rm\Omega$ \cite{e-squid} and FDM LC's are designed for the frequency resonances between 1 MHz and 3.8 MHz. Parasitics capacitors at SQUID summing point have empirical values of 150 pF and 850 pF (next to the right red box in Fig.\ref{fig1}). Further details of the setup can be found in \cite{ref10}, \cite{ref3}, \cite{ref4}. In the simulation demonstrated in Fig.\ref{fig1} the impact of cryogenic harness and input coil of the SQUID has been examined. 
 Position of the snubber in use is shown in the schematics in Fig.~\ref{fig1} at Summing Point A. The snubber circuit in use has the R 2.2 $\rm\Omega$ and C 10 nF which are the optimized values to suppress the OBR based on the LT-SPICE simulation. As can be seen in the simulation results in Fig.~\ref{fig2}, the OBR resonance at 98.7 MHz caused by the parasitic capacitances together with the input coil of SQUID, is dampened by the snubber in use at the input of the first stage SQUID. The peak at 15.9 MHz is related to the LC resonance made by the characteristic impedance of cryogenic harness and could be dampened by using the 2$^{\rm nd}$ snubber (only in simulation, not in use) at the AC bias line before $R_{\rm shunt}$.   
\\ \\ \\
Next, we simulated the entire BBFB FDM readout using LT-SPICE software. The block diagram of the FDM readout shown in Fig.~\ref{fig3} is actually a simplified version of the readout system in Fig.~\ref{fig01}. It includes the AC bias circuit, harness, two-stage SQUID, Warm Front End Electronics (WFEE), BBFB resistor and an external LNA. For the SQUIDs, we used small signal model where the SQUID amplification is linear. This approximation is accurate enough for the BBFB FDM readout. For advanced SPICE modelling of the DC SQUID see \cite{refSQ1},\cite{refSQ2}. Only the snubber in use at summing point is included in the simulation. The results of simulation are shown in Fig.~\ref{fig4}. Comparing the results with those in Fig.~\ref{fig2}, it can be seen that by applying the BBFB resistor and snubber, the OBR peaks have been significantly reduced and their magnitude is well below the magnitude of the FDM signal (1 MHz-3.8 MHz). It also shows the impact of long harness (few meters for flight model) which generates repetitive OBR spikes at higher frequenies (peaks appear at frequencies higher than 50 MHz). For an AC bias of 1 V, the FDM output level is around 20 dB and the Max. OBR peak is ~-10 dB. Although in simulation, the magnitude of the OBR spikes is well below the FDM signal, it should be noted that nearby OBRs could in principle destabilize the SQUID bias, increase the irregularities in the V-$\rm\Phi$ curve and increase the readout noise\cite{ref32}.   
\section{Measurement}
\begin{figure}[h]
\centering
\includegraphics[width=8cm]{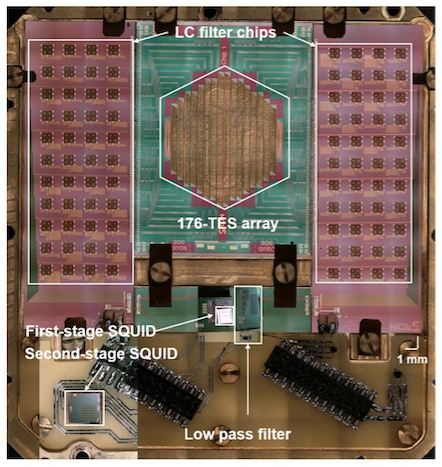}   
\caption{\label{fig02}
A photo of a prototype 176 TES array developed at SRON, after \cite{ref10}. The chip includes LC resonators, first and second stage SQUID and the snubber circuit(RC filter ). 
}
\end{figure}
\begin{figure}[h]

\includegraphics[width=13.5cm]{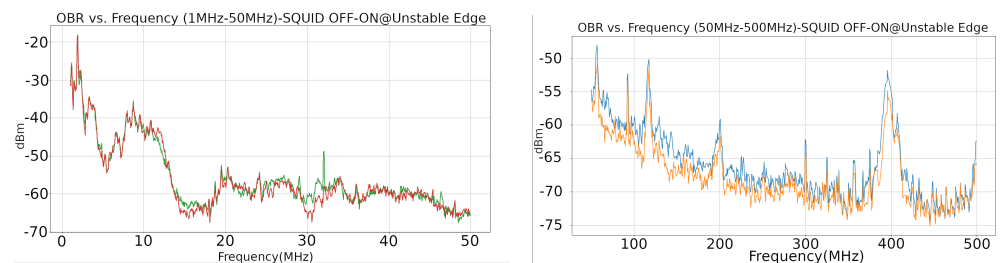}


\caption{\label{fig7}
A comparison of the OBR level after amplification by an external LNA when SQUID is OFF-ON: Left, 1 MHz-50 MHz with the SQUID OFF(Red)-ON (Green) Right, 50 MHz-500 MHz with the SQUID OFF(Orange)-ON (Blue). 
}
\end{figure}

Using an external broadband LNA at room temperature, we examined the OBR level at a broad frequency range from 1MHz up to 500 MHz. The FDM setup including the TES array and two-stage SQUID is inside a cryostat at 50mK and 4K regimes. In Fig.~\ref{fig02}, a photo of 176 TES array developed at SRON is shown where the LC resonators, SQUIDs and the snubber(Low-Pass RC filter) are identified with white boxes. 
Test has been repeated for various SQUID DC bias  (two edges)  and  SQUID is turned On and Off.  \\
To optimize the noise performance of the external LNA, we repeated the test using a broadband (1 MHz-2 GHz)
Cryo-LNA at 77K (using liquid nitrogen) for 2 seperate FDM setups at SRON.   
Selected results are shown in Fig.~\ref{fig7} (1 MHz-50 MHz, left and 50 MHz-500 MHz, right).  It can be seen that the SQUID has negligible impact on the OBR. The measurement with Cryo-LNA did not change the results either. It is concluded that
the observed OBR is either due to the harness impedance mismatch or due to the pickup noise at the lab. It should be noted that in the current prototype setup, the FEE amplifier, which amplifies the signal before the ADC, is at room temperature. 
 The results in Fig.~\ref{fig7}which are measured using an external LNA only shows the impact of the SQUID on the OBR. To see the modulation of the FDM readout vs the OBR, we also took data from the output of the FEE while the snubber was in use. Similar to results in Fig.~\ref{fig7}, several resonances occur at frequencies below 10 MHz suggesting that nearby OBR could be due to the harmonics or intermodulation products of the FDM resonances. 


\section{Conclusion}
Here is a summary of the results: 
\begin{description}
\item[-] Simulation shows that the long harness for flight model could generate repetitive spikes in the spectrum. The important factor to consider is the characteristic impedance of the harness (e.g. twisted pairs) which should be matched to the output of SQUID.
\item[-] In addition to the harness, an OBR peak could occur due to the capacitances at the input coil of the SQUID. The capacitors (150 pF,850 pF in our model) represent  the parasitic junction capacitances which are well known in the literature(e.g. \cite{ref7}, \cite{ref8}).
\item[-] There are also inductance of the co-planar and bonding wires as well as inductance of microstrip lines but these inductances are negligible compared to the inductors used in the LC filter chip \cite{ref9}.   
\item[-] Implementing a low-pass RC circuit (snubber) at the input of SQUID dampened the OBR so the first peak observed around 20 MHz which is a safe margin for the 4 MHz FDM in use in the prototype setup.
\item[-] OBRs are more intense at 50 mK compared to those at 4K and seem to be caused by harness impedance mismatch rather than the SQUID as the data measured with the SQUID On and Off has no significant changes.
\item[-] A 2nd snubber at the AC bias line before $R_{shunt}$ could in principle further suppress the OBR. This is not implemented in the setup but could be investigated for future study.\\
\item [-] In addition to the use of snubber (low pass RC filter), placing the RF amplifier at a shielded cryostat and proper filtering after amplification could improve the FDM performance and reduce the OBR.
\end{description}

\begin{acknowledgements}
 \enspace
  We thank Kevin Ravensberg, Marcel van Litsenburg and Alex Simon at SRON workshop. 
We also thank anonymous referees for their constructive comments. SRON is
supported financially by NWO, the Netherlands Organization
for Scientific Research. Amin Aminaei is the recipient of the Brinson Prize Fellowship at UC Davis. 
\end{acknowledgements}

\end{document}